\documentclass[conference]{IEEEtran}
\usepackage{epsfig,cite,stfloats,bm,amssymb,amsmath,color,subfigure,graphics}

\long\def\symbolfootnote[#1]#2{\begingroup
\def\thefootnote{\fnsymbol{footnote}}
\footnote[#1]{#2}\endgroup}

\begin{document}
\title{Pilot Decontamination via PDP Alignment}

\author{\IEEEauthorblockN{Xiliang Luo$^1$, Xiaoyu Zhang$^1$, Hua Qian$^2$, and Kai Kang$^2$}
\IEEEauthorblockA{$^1$ School of Information Science and Technology, ShanghaiTech
University, Shanghai, China\\
$^2$ Shanghai Advanced Research Institute, Chinese Academy of Sciences, Shanghai, China\\
Email: luoxl@shanghaitech.edu.cn}}

\maketitle

\symbolfootnote[0]{This work was supported through the startup fund from ShanghaiTech University under the grant no. F-0203-14-008 and was supported in part
by the National Key Science and Technology 863 Project under Grant No. 2015AA01A709
and the Science and Technology Commission Foundation of Shanghai under Gant No.
14511100200.}

\begin{abstract}
In this paper, we look into the issue of intra-cell uplink (UL) pilot
orthogonalization and schemes for mitigating the inter-cell pilot contamination
with a realistic massive multi-input multi-output (MIMO) orthogonal
frequency-division multiplexing (OFDM) system model. First, we show how to align
the power-delay profiles (PDP) of different users served by one BS so that the
pilots sent within one common OFDM symbol are orthogonal. From the derived
aligning rule, we see much more users can be sounded in the same OFDM symbol as
their channels are sparse in time. Second, in the case of massive MIMO, we show
how PDP alignment can help to alleviate the pilot contamination due to inter-cell
interference. We demonstrate that, by utilizing the fact that different paths in
time are associated with different angles of arrival (AoA), the pilot
contamination can be significantly reduced through aligning the PDPs of the users
served by different BSs appropriately. Computer simulations further convince us
PDP aligning can serve as the new baseline design philosophy for the UL pilots
in massive MIMO.
\end{abstract}


\section{Introduction}

% Massive MIMO overview
Massive multiple-input multiple-output (MIMO) has been regarded as one
of the enabling technologies in next generation wireless
communications\cite{marzetta10twc, resek13spmag,larsson14commag,lu14jstsp}. Considering the large
number of antennas at the base station (BS), it becomes almost imperative to
rely on the time-division duplexing (TDD) channel reciprocity to learn the downlink
(DL) channel state information (CSI) from the uplink (UL) channel measurements at
the BS to avoid the huge overhead for CSI feedback as in frequency-division
duplexing (FDD) systems. Even though the antenna array size is large at the BS,
each user is only equipped with a few antennas, e.g. only one antenna. A user
just needs to send one UL pilot sequence per transmit antenna to facilitate
the BSs to acquire reliable estimates of the corresponding many UL channels
from the user to the large antenna array at the BS.

To ensure best CSI estimation quality, we want to allocate orthogonal UL pilot
sequences to different users so that the pilot transmissions do not interfere with
each other. But in reality, within a limited time period and a limited bandwidth,
there are only a limited number of orthogonal pilot sequences. As the number of
users becomes large, non-orthogonal pilot sequences are re-used by the users
served by different BSs, which gives rise to the so-called pilot contamination
\cite{marzetta10twc,lu14jstsp} and is one limiting factor in multi-cell massive
MIMO systems.

Various approaches have been proposed to alleviate the pilot contamination issue in
massive MIMO. Recent works include \cite{yin2013jsac,yin2014jstsp,yin2015spawc,hu16twc,muller14jstsp}. See also \cite{lu14jstsp} and
references therein for a brief overview.
In \cite{yin2013jsac} and \cite{yin2014jstsp}, pilot decontamination was achieved
by utilizing the fact that users with non-overlapping angles of arrival (AoA) enjoy
asymptotic orthogonal covariance matrices. In \cite{yin2015spawc}, AoA diversity
and amplitude-based projection were jointly exploited to null the
pilot contamination and achieve better channel estimation. A blind singular value decomposition (SVD) based method was proposed in \cite{muller14jstsp} to separate the signal subspace from the interference subspace. In \cite{hu16twc},
least-squares (LS) channel estimate was derived by treating the blindly detected UL
data as pilot symbols and the pilot contamination effect was shown to diminish as
the data length grew.

Unlike previous studies where a single narrow-band channel was typically
assumed, in this paper, we look into the issue of intra-cell pilot orthogonalization
and schemes for mitigating the inter-cell pilot contamination with a realistic massive MIMO OFDM system model. First, we examine how to align the power-delay
profiles (PDP, a.k.a. delay power spectrum in \cite{DigitalComBook}) of different
users served by one BS so that the pilots sent within one common OFDM symbol are
orthogonal. From the aligning rule, we can see much more users can be sounded in the
same OFDM symbol as their channels are sparse in time. In the case of massive MIMO,
to alleviate the pilot contamination when the schemes in \cite{yin2013jsac,yin2014jstsp,yin2015spawc} do not apply well due to interfering users' overlapping AoAs, we further propose to exploit the fact that different paths
in time are associated with different AoAs and the pilot contamination can be significantly reduced by aligning the PDPs of the users served by different BSs
appropriately. Thus, PDP aligning can serve as the new baseline design philosophy
for massive MIMO UL pilots.

% Structure
This paper is organized as follows: Section \ref{SecSysModel} describes the massive
MIMO OFDM system model and the channel model. Section \ref{OrthoPilot} provides a
sufficient condition for orthogonal pilots design through PDP aligning, which is
also applicable to conventional MIMO systems. Then we explain how to mitigate the
pilot interference in the case of massive MIMO by PDP alignment in Section
\ref{IntfReduction}. Low-complexity pilot designs are provided in Section
\ref{Protocol}. Corroborating simulation results are provided in Section
\ref{Performance} and Section \ref{conclusion} concludes the paper.

{{\it Notations:} ${\sf Diag}\{\cdots\}$ denotes the diagonal matrix with diagonal
elements defined inside the curly brackets. ${A}(i,j)$ refers to the $(i,j)$th
entry of matrix $\bm A$ and $a(i)$ stands for the $i$-th entry of the vector ${\bm a}$. $\bm I_N$ denotes the $N\times N$ identity matrix. ${\sf E}[\cdot]$, ${\tt Tr}(\cdot)$, $(\cdot)^{\dagger}$, $(\cdot)^{T}$, and $(\cdot)^*$ represent
expectation, matrix trace, Hermitian operation, transpose, and conjugate operation respectively.}

\section{System Model and Channel Model}\label{SecSysModel}

We consider an MIMO-OFDM system with $B$ macro BSs. Each BS is equipped with
a massive antenna array of size $M$ and serves $K$ single-antenna
users. Regarding the OFDM waveform, we adopt the following notations:
\begin{itemize}
  \item $N$: total number of sub-carriers, a.k.a. tones;
  \item $T$: time duration of one OFDM symbol;
  \item $T_c:=T/N$: time duration of one chip;
  \item $N_{cp}$: cyclic prefix length in $T_c$.
\end{itemize}

As the delay spread of each user's channel is less than $N_{cp}$, after standard
OFDM receiver processing, the received signal at the $m$-th antenna in BS-$b$
can be expressed as
\begin{eqnarray}
{\bm y}_m^{(b)}=\sum_{l=1}^{B}\sum_{k=1}^{K}\sqrt{\rho_k^{(l)}}{\bm S}_k^{(l)}{\bm H}_{k,m}^{(l,b)}+{\bm w}_m^{(b)},\label{SysModel}
\end{eqnarray}
where ${\bm y}_m^{(b)}$ is an $N\times1$ vector containing the received signal over all the tones and the summation is taken over all the BSs and all the served users.
For user-$k$ in BS-$l$ (we will denote it as user-$(l,k)$ in the sequel),
$\rho_k^{(l)}$ denotes the transmitted power over
each tone and ${\bm S}_k^{(l)}$ is an $N\times N$ diagonal matrix with diagonal
entries being the transmitted pilot sequence. The frequency response of the channel
from user-$(l,k)$ to the $m$-th antenna at BS-$b$ is ${\bm H}_{k,m}^{(l,b)}$ and
${\bm w}_m^{(b)}$ represents the receiver noise with covariance ${\sf E}[{\bm w}_m^{(b)}{\bm w}_m^{(b)\dagger}]=\sigma^2 {\bm I}$.

The channel frequency response is the discrete Fourier transform (DFT) of
the corresponding channel impulse response (CIR) in time domain, i.e.
\begin{equation}
{\bm H}_{k,m}^{(l,b)}={\bm F}{\bm h}_{k,m}^{(l,b)}, \label{CFR_CIR}
\end{equation}
where ${\bm F}$ stands for the unitary FFT matrix defined as ${F}(k,n)=\exp\{-j2\pi (k-1)(n-1)/N\}/\sqrt{N}$ and ${\bm h}_{k,m}^{(l,b)}$ is the CIR
with the following PDP:
\begin{equation}
{\bm P}_k^{(l,b)}:={\sf E}[{\bm h}_{k,m}^{(l,b)}{\bm h}_{k,m}^{(l,b)\dagger}],
\end{equation}
where we have assumed the channels from one user to the antennas at one BS share a
common PDP. Assuming uncorrelated scattering as in \cite{DigitalComBook}, i.e. the
scattering at two different paths is uncorrelated, the PDP matrix ${\bm P}_k^{(l,b)}$ becomes diagonal. Combining (\ref{SysModel}) and (\ref{CFR_CIR}), we
have
\begin{eqnarray}
{\bm y}_m^{(b)}=\sum_{l=1}^{B}\sum_{k=1}^{K}\sqrt{\rho_k^{(l)}}{\bm S}_k^{(l)}{\bm F}{\bm h}_{k,m}^{(l,b)}+{\bm w}_m^{(b)}.\label{SysModel2}
\end{eqnarray}

\begin{figure}[t]
\centering \epsfig{file=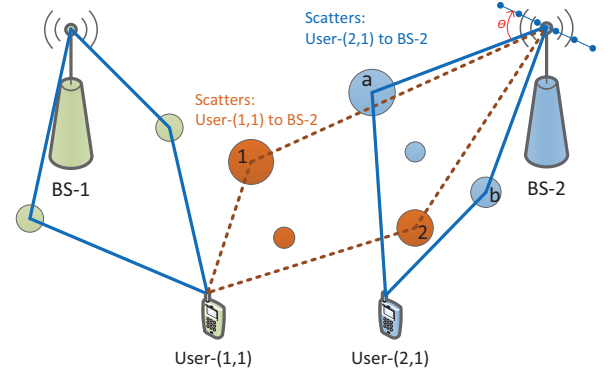,width=0.43\textwidth}
\caption{Spatial channel model.} \label{ChModelFig}
\vspace{-0.3cm}
\end{figure}

In order to characterize the spatial covariance among multiple receive antennas at
the BSs, we adopt the multi-path spatial channel model (SCM) defined in \cite{SCM}
and \cite{cost2100}, which is illustrated in Fig. \ref{ChModelFig}. Each resolvable
channel path corresponds to one independent scatterer including $Q$ sub-paths:
\begin{equation}
\left[h_{k,1}^{(l,b)}(n),...,h_{k,M}^{(l,b)}(n)\right]^T= \sqrt{\frac{P_{k}^{l,b}(n,n)}{Q}}\sum_{q=1}^Q {\bm a}(\theta_{k,n,q}^{(l,b)})e^{j\phi_q},\label{channelModel}
\end{equation}
where the arriving angles: $\{\theta_{k,n,q}^{(l,b)}\}_{q=1}^{Q}$ of the sub-paths
are uniformly distributed within the angle spread (AS) of this path, the phases: $\{\phi_q\}$ are drawn from a uniform distribution over $[0,2\pi]$,
and the vector ${\bm a}(\theta)$ stands for the steering vector of the receive
antenna array when the AoA of the incoming path is $\theta$. Assuming a uniform
linear array (ULA), the steering vector can be expressed as:
\begin{equation}
{\bm a}(\theta)=\left[1, e^{-j2\pi D/\lambda\cos(\theta)},..., e^{-j2\pi(M-1)D/\lambda\cos(\theta)}\right]^T,
\end{equation}
where $D$ is the antenna spacing and $\lambda$ denotes the carrier wavelength.
As $M\rightarrow\infty$, we can obtain one noticeable result as follows: $\forall \theta_1\neq \theta_2\in(0,\pi)$, as $D<\lambda/2$,
\begin{eqnarray}
&&\lim_{M\rightarrow\infty}\frac{{\bm a}(\theta_1)^{\dagger}{\bm a}(\theta_2)}{\sqrt{{\bm a}(\theta_1)^{\dagger}{\bm a}(\theta_1)}\sqrt{{\bm a}(\theta_2)^{\dagger}{\bm a}(\theta_2)}}=\nonumber\\
&&\lim_{M\rightarrow\infty}\left|\frac{\sin(M\pi D(\cos(\theta_1)-\cos(\theta_2))/\lambda)}{M\sin(\pi D(\cos(\theta_1)-\cos(\theta_2))/\lambda)}\right|\leq\nonumber\\
&&\lim_{M\rightarrow\infty}\left|\frac{1}{M\sin(\pi D(\cos(\theta_1)-\cos(\theta_2))/\lambda)}\right|=0, \label{AsympOrtho}
\end{eqnarray}
where we have assumed $\theta_1$ and $\theta_2$ are independent of $M$.
This result indicates the asymptotic orthogonality of the paths
arriving from different angles. In the following sections, we will take
advantage of this important fact to design UL pilots.

\section{Orthogonal Designs via PDP Aligning} \label{OrthoPilot}
Using the observation in (\ref{SysModel2}), we can obtain the MMSE estimate
for the channel between user-$(b,u)$ and the $m$-th antenna at BS-$b$ as follows:
\begin{eqnarray}
&&\hspace{-0.8cm}\hat{{\bm h}}_{u,m}^{(b,b)}={\sf E}[{\bm h}_{u,m}^{(b,b)}{\bm y}_m^{(b)\dagger}]\left({\sf E}[{\bm y}_m^{(b)}{\bm y}_m^{(b)\dagger}]\right)^{-1}{\bm y}_m^{(b)}\nonumber\\
&&\hspace{-0.8cm}=\sqrt{\rho_{u}^{(b)}}{\bm P}_{u}^{(b,b)}{\bm F}^{\dagger}{\bm S}_u^{(b)\dagger}\cdot\nonumber\\
&&\hspace{-0.8cm}\left(\sigma^2{\bm I}+ \rho_{u}^{(b)}{\bm S}_u^{(b)}{\bm F}{\bm P}_{u}^{(b,b)}{\bm F}^{\dagger}{\bm S}_u^{(b)\dagger}+\Delta_1+\Delta_2\right)^{-1}{\bm y}_m^{(b)},
\end{eqnarray}
where
$$\Delta_1:=\sum_{k=1,k\neq u}^{K}\rho_{k}^{(b)}{\bm S}_k^{(b)}{\bm F}{\bm P}_{k}^{(b,b)}{\bm F}^{\dagger}{\bm S}_k^{(b)\dagger}$$ contains the interference
from the intra-cell users, and
$$\Delta_2:=\sum_{l=1,l\neq b}^{B}\sum_{k=1}^{K}\rho_{k}^{(l)}{\bm S}_k^{(l)}{\bm F}{\bm P}_{k}^{(l,b)}{\bm F}^{\dagger}{\bm S}_k^{(l)\dagger}$$ includes the
inter-cell interference. Note in the above derivation, we have assumed the channels
among different users are independent.

Defining the channel estimation error
as ${\bm \epsilon}:={{\bm h}}_{u,m}^{(b,b)}-\hat{{\bm h}}_{u,m}^{(b,b)}$, we can obtain its covariance as follows:
\begin{eqnarray}
{\sf E}[{\bm \epsilon}{\bm \epsilon}^{\dagger}]\hspace{-0.3cm}&=&\hspace{-0.3cm}
{\bm P}_u^{(b,b)}-{\rho_{u}^{(b)}}{\bm P}_{u}^{(b,b)}{\bm F}^{\dagger}{\bm S}_u^{(b)\dagger}\cdot\nonumber\\
&&\hspace{-0.3cm}\left( \sigma^2{\bm I}+ \rho_{u}^{(b)}{\bm S}_u^{(b)}{\bm F}{\bm P}_{u}^{(b,b)}{\bm F}^{\dagger}{\bm S}_u^{(b)\dagger}+\Delta_1+\Delta_2\right)^{-1}\cdot\nonumber\\
&&\hspace{-0.3cm}{\bm S}_u^{(b)}{\bm F}{\bm P}_{u}^{(b,b)}. \label{MSE}
\end{eqnarray}
Without loss of generality, we let ${\bm S}_u^{(b)}{\bm S}_u^{(b)\dagger}={\bm I}$,
i.e. the pilot sequence enjoys constant unit modulus. Then, we can rewrite (\ref{MSE}) as
\begin{eqnarray}
{\sf E}[{\bm \epsilon}{\bm \epsilon}^{\dagger}]\hspace{-0.3cm}&=&\hspace{-0.3cm}
{\bm P}_u^{(b,b)}-{\rho_{u}^{(b)}}{\bm P}_{u}^{(b,b)}\cdot\nonumber\\
&&\hspace{-0.3cm}\left( \sigma^2{\bm I}+ \rho_{u}^{(b)}{\bm P}_{u}^{(b,b)}+\tilde{\Delta}_1+\tilde{\Delta}_2\right)^{-1}{\bm P}_{u}^{(b,b)}, \label{MSE2}
\end{eqnarray}
where $\tilde{\bm \Delta}_1={\bm F}^{\dagger}{\bm S}_u^{(b)\dagger}{\bm \Delta}_1{\bm S}_u^{(b)}{\bm F}$ and $\tilde{\bm \Delta}_2={\bm F}^{\dagger}{\bm S}_u^{(b)\dagger}{\bm \Delta}_2{\bm S}_u^{(b)}{\bm F}$.
In the absence of interference, the corresponding MSE covariance can be computed
accordingly as
\begin{equation}
{\sf E}[{\bm \epsilon_0}{\bm \epsilon}_0^{\dagger}]=
{\bm P}_u^{(b,b)}-{\rho_{u}^{(b)}}{\bm P}_{u}^{(b,b)}
\left(\sigma^2{\bm I}+ \rho_{u}^{(b)}{\bm P}_{u}^{(b,b)}\right)^{-1}{\bm P}_{u}^{(b,b)}.\label{MSE3_clean}
\end{equation}
To achieve the orthogonality between the received pilots from the interfering
user-$(l,k)$ and the targeted user-$(b,u)$, from (\ref{MSE2}) and
(\ref{MSE3_clean}), we need to ensure ${\bm S}_u^{(b)}$ and
${\bm S}_k^{(l)}$ satisfy the following condition:
\begin{eqnarray}
&&\hspace{-1.0cm}{\bm P}_{u}^{(b,b)}\left(\sigma^2{\bm I}+ \rho_{u}^{(b)}{\bm P}_{u}^{(b,b)}\right)^{-1}{\bm P}_{u}^{(b,b)}=\nonumber\\
&&\hspace{-1.0cm}{\bm P}_{u}^{(b,b)}
\left(\sigma^2{\bm I}+ \rho_{u}^{(b)}{\bm P}_{u}^{(b,b)}
+\rho_{k}^{(l)}{\bm \Theta}{\bm P}_{k}^{(l,b)}{\bm \Theta}^{\dagger}\right)^{-1}{\bm P}_{u}^{(b,b)},\label{OrthoCond1}
\end{eqnarray}
where ${\bm \Theta}:={\bm F}^{\dagger}{\bm S}_u^{(b)\dagger}{\bm S}_k^{(l)}{\bm F}$.
From (\ref{OrthoCond1}), we can establish the following requirement on the pilot
sequences:\\
\noindent{\bf Proposition 1:}
{\it To achieve orthogonality between the UL pilots from user-$(l,k)$ and
user-$(b,u)$ at each receive antenna in the BS, the constant unit-modulus pilot
sequences need to satisfy the following condition:
\begin{eqnarray}
{\bm P}_{u}^{(b,b)}{\bm \Theta}{\bm P}_{k}^{(l,b)}{\bm \Theta}^{\dagger}={\bm 0}.\label{OrthoCond2}
\end{eqnarray} }

In the following discussions, we will assume the following pilot sequences as in the
LTE UL \cite{LTEBook}:
\begin{equation}
{\bm S}_k^{(l)} = {\sf Diag}\left\{1,e^{j\frac{2\pi\tau_{l,k}}{N}},...,e^{j\frac{2\pi\tau_{l,k}(N-1)}{N}}
\right\}\cdot{\bm S}_0, \label{CyclicShift}
\end{equation}
where $\tau_{l,k}$ is the amount of cyclic time shifts and ${\bm S}_0$ is the base
unshifted sequence with constant modulus. With the above sequence designs, the
matrix ${\bm \Theta}$ becomes unitary and circulant with the first column vector
taking the following form:
\begin{eqnarray}
{\bm \Theta}(:,1)^T=[\underbrace{0,...,0}_{N-\Delta\tau},1,\underbrace{0,...,0}_{\Delta\tau-1}],\label{FirstColumn}
\end{eqnarray}
where $\Delta\tau:=\tau_{l,k}-\tau_{b,u}$ refers to the amount of relative cyclic
shifts between the two users. From (\ref{OrthoCond2}), we obtain the
corresponding requirement on $\Delta\tau$ as:
\begin{equation}
{\bm P}_{u}^{(b,b)}\tilde{\bm P}_{k}^{(l,b)}={\bm 0},\label{OrthoCond3}
\end{equation}
where $\tilde{\bm P}_{k}^{(l,b)}:={\bm \Theta}{\bm P}_{k}^{(l,b)}{\bm \Theta}^{\dagger}$ is the result of cyclicly shifting the diagonals of
${\bm P}_{k}^{(l,b)}$ by $-\Delta\tau$. The orthogonality condition in (\ref{OrthoCond3}) simply states that, to
ensure orthogonal pilots between two users, the amount of the relative cyclic shifts between the two users should be chosen such that the supports of the shifted PDPs
are non-overlapping.

\begin{figure}
\centering \epsfig{file=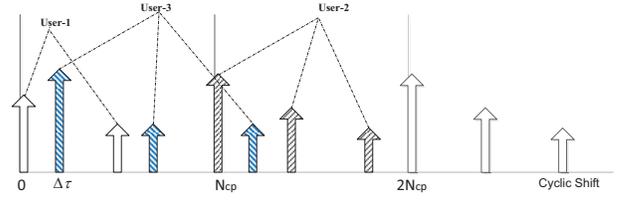,width=0.45\textwidth}
\caption{Orthogonality via PDP alignment.} \label{PDPAlign1}
\vspace{-0.4cm}
\end{figure}

Conventional designs assume that the PDPs of all users are confined within
the first $N_{cp}$ taps \cite{LTEBook}. In order to ensure orthogonality in (\ref{OrthoCond3}), we see up to $\frac{N}{N_{cp}}$ users\footnote{Here we assume
$N_{cp}$ divides $N$. Otherwise, we can do flooring: $\lfloor \frac{N}{N_{cp}}\rfloor$.}  can transmit pilots in the same
OFDM symbol and the relative cyclic shift values among users are:
$\Delta\tau=kN_{cp},k=0,1,...,N/N_{cp}-1$. In fact,
from the general condition specified in (\ref{OrthoCond3}), we can easily see that
the amount of cyclic shifts required can be well below $N_{cp}$ when the PDPs are sparse in time (see also Fig. \ref{PDPAlign1}).

Even though we can exploit the channel sparsity and the proposed PDP alignment as
illustrated in Fig. \ref{PDPAlign1}, it is clear that there are only a limited
number of orthogonal pilot sequences available within the channel coherence time.
Pilot contamination refers to the fact that, as $B\cdot K$ becomes large, we can
not provide each user an orthogonal pilot sequence. Instead, a typical solution is
to orthogonalize the users served by a common BS while allowing non-orthogonal pilot
sequences among users served by different BSs. In the next section, we will
examine a novel scheme applicable in the case of massive MIMO to reduce the
inter-cell interference.

\section{PDP Alignment for Interference Reduction} \label{IntfReduction}

Assume the users $\{1,...,K_l\}$ served by BS-$l$ are assigned the
cyclic shifts $\{\tau_{l,1},...,\tau_{l,K_l}\}$ in
(\ref{CyclicShift}) and the orthogonality
condition in (\ref{OrthoCond3}) is met with these cyclic shifts.
Since all the orthogonal cyclic shift resources are used
to enable intra-cell orthogonality, the users served by different
BSs have to reuse the same set of pilot sequences. From (\ref{SysModel2}),
we can have the following signal model at the $m$-th antenna of BS-$b$:
\begin{eqnarray}
{\bm y}_m^{(b)}&=&\sum_{l=1}^{B}\sum_{k=1}^{K_{l}}\sqrt{\rho_k^{(l)}}{\bm S}_k^{(l)}{\bm F}{\bm h}_{k,m}^{(l,b)}+{\bm w}_m^{(b)}.\label{SysModel3}
\end{eqnarray}
Then, we can obtain
\begin{eqnarray}
{\bm z}_m^{(b)}&:=&{\bm F}^{\dagger}{\bm S}_0^{\dagger}{\bm y}_m^{(b)}=\sum_{l=1}^{B}\sum_{k=1}^{K_l}\sqrt{\rho_k^{(l)}}{\bm \Theta}_k^{(l)}{\bm h}_{k,m}^{(l,b)}+{\bm \omega}_m^{(b)}\nonumber\\
&=&\sum_{l=1}^{B}\breve{\bm h}_{m}^{(l,b)}+{\bm \omega}_m^{(b)},\label{SysModel4}
\end{eqnarray}
where $\breve{\bm h}_{m}^{(l,b)}:=\sum_{k=1}^{K_{l}}\sqrt{\rho_k^{(l)}}{\bm \Theta}_k^{(l)}{\bm h}_{k,m}^{(l,b)}$ is the aggregate channel of all those $K_l$
non-interfering intra-cell orthogonal users served by BS-$l$, ${\bm \Theta}_k^{(l)}:={\bm F}^{\dagger}{\bm S}_0^{\dagger}{\bm S}_k^{(l)}{\bm F}$ is a
circulant matrix with the first column vector defined as in (\ref{FirstColumn}) with $\Delta\tau=\tau_{l,k}$, and the noise term ${\bm\omega}_m^{(b)}$ has the same covariance as ${\bm w}_m^{(b)}$. Stacking
the $n$-th taps of $\{{\bm z}_m^{(b)}\}_{m=1}^M$ into an $M\times1$ vector as:
${\bm g}_n^{(b)}:=[{z}_1^{(b)}(n),...,{z}_M^{(b)}(n)]^T $, we can get
\begin{eqnarray}
{\bm g}_n^{(b)}={\bm h}_n^{(b,b)}+
\sum_{l=1,l\neq b}^{B}{\bm h}_{n}^{(l,b)}+{\bm \omega}^{(b)}, \label{SysModel5}
\end{eqnarray}
where ${\bm h}_n^{(l,b)}$ denotes the vector of the $n$-th taps in the aggregate
channels from the orthogonal users served by BS-$l$ to BS-$b$, i.e. ${\bm h}_n^{(l,b)}:=[\breve{h}_{1}^{(l,b)}(n),...,\breve{h}_{M}^{(l,b)}(n)]^T$ and ${\bm \omega}^{(b)}:=[{\omega}_{1}^{(b)}(n),...,{\omega}_{M}^{(b)}(n)]^T $.
Since the circulant matrix ${\bm \Theta}_k^{(l)}$ carries out cyclic shift
operation, we have
\begin{eqnarray}
\breve{h}_{m}^{(l,b)}(n)=\sum_{k=1}^{K_l}\sqrt{\rho_k^{(l)}}
{h}_{k,m}^{(l,b)}(n+\tau_{l,k}).\label{aggTap}
\end{eqnarray}
Note that for channels of limited delay spread, only a few users
will have non-zero contribution toward the aggregated channel tap in (\ref{aggTap}).

Denoting the spatial covariance matrix of the $n$-th taps in the aggregate channels
from users in BS-$l$ to BS-$b$ across the $M$ BS antennas as ${\bm C}_{n}^{(l,b)}$,
i.e. ${\bm C}_{n}^{(l,b)}:={\sf E}\left[{\bm h}_{n}^{(l,b)}{\bm h}_{n}^{(l,b)\dagger}\right]$, with the signal model in (\ref{SysModel5}), we can derive the MMSE estimate of the desired
channel ${\bm h}_n^{(b,b)}$ as:
\begin{eqnarray}
&&\hspace{-0.9cm}\hat{\bm h}_n^{(b,b)}={\sf E}[{\bm h}_n^{(b)}{\bm g}_n^{(b)\dagger}]({\sf E}[{\bm g}_n^{(b)}{\bm g}_n^{(b)\dagger}])^{-1}{\bm g}_n^{(b)}\nonumber\\
&&\hspace{-0.9cm}={\bm C}_{n}^{(b,b)}\cdot\left(\sigma^2{\bm I}_{M}+ \sum_{l=1}^{B}{\bm C}_{n}^{(l,b)}   \right)^{-1}\hspace{-0.3cm}{\bm g}_n^{(b)}.
\end{eqnarray}
The covariance of the estimation error vector: ${\bm \epsilon}_{n}^{(b,b)}:={\bm h}_n^{(b,b)}-\hat{\bm h}_n^{(b,b)}$ can be found as follows:
\begin{eqnarray}
&&\hspace{-0.8cm}{\sf E}[{\bm \epsilon}_{n}^{(b,b)}{\bm \epsilon}_{n}^{(b,b)\dagger}] =
{\bm C}_{n}^{(b,b)}- \nonumber\\
&&\hspace{-0.8cm}{\bm C}_{n}^{(b,b)}\left(\sigma^2{\bm I}_{M}+{\bm C}_{n}^{(b,b)}+\sum_{l=1,l\neq b}^B{\bm C}_{n}^{(l,b)} \right)^{-1}\hspace{-0.1cm}{\bm C}_{n}^{(b,b)}. \label{TapMSE}
\end{eqnarray}
Let $\sum_{l=1,l\neq b}^B{\bm C}_{n}^{(l,b)}={\bm U}{\bm \Sigma}{\bm U}^{\dagger}$ and ${\bm C}_{n}^{(b,b)}={\bm V}{\bm \Lambda}{\bm V}^{\dagger}$, where
$\bm U$ ($\bm V$) is an $M\times r$ ($M\times r'$) matrix consisting of $r$ ($r'$) eigenvectors and $\bm \Sigma$ ($\bm \Lambda$)
is an $r\times r$ ($r'\times r'$) diagonal matrix consisting of $r$ ($r'$) non-zero eigenvalues. From (\ref{TapMSE}), we get
\begin{eqnarray}
&&\hspace{-1cm}{\sf E}[{\bm \epsilon}_{n}^{(b,b)}{\bm \epsilon}_{n}^{(b,b)\dagger}] = {\bm C}_{n}^{(b,b)}-\nonumber\\
&&\hspace{-0.9cm}{\bm C}_{n}^{(b,b)}\left(\sigma^2{\bm I}_{M}+{\bm C}_{n}^{(b,b)}\right)^{-1}\hspace{-0.1cm}{\bm C}_{n}^{(b,b)} + {\bm R}_{n}^{(b)}, \label{TapMSE2}
\end{eqnarray}
where the residual matrix ${\bm R}_{n}^{(b)}$ is of the following form
\begin{eqnarray}
{\bm R}_{n}^{(b)}&=&\left(\sigma^2{\bm I}_{M}+{\bm C}_{n}^{(b,b)}\right)^{-1}{\bm V}{\bm \Lambda}{\bm V}^{\dagger}{\bm U}\cdot\nonumber\\
&&\left({\bm \Sigma}^{-1}+{\bm U}^{\dagger} \left(\sigma^2{\bm I}_{M}+{\bm C}_{n}^{(b,b)}\right)^{-1} {\bm U} \right)^{-1}\cdot\nonumber\\
&&{\bm U}^{\dagger}{\bm V}{\bm \Lambda}{\bm V}^{\dagger} \left(\sigma^2{\bm I}_{M}+{\bm C}_{n}^{(b,b)}\right)^{-1}. \label{ResidualR}
\end{eqnarray}
To obtain a reliable estimate of ${\bm h}_n^{(b,b)}$, we would
like to make the subspaces spanned by the interference term and the signal term as orthogonal as possible. When ${\sf span}\{\bm U\}\perp {\sf span}\{\bm V\}$, we
have ${\bm R}_{n}^{(b)}={\bm 0}$ and achieve the interference-free estimation
performance.

Channel taps (a.k.a. paths) of different time delays are originating
from different scattering clusters. Similar to (\ref{AsympOrtho}),
when the angles of arrival (AoA) of two paths do not overlap, it has been
shown that the associated covariance matrices span orthogonal
subspaces asymptotically as $M\rightarrow\infty$ \cite{yin2013jsac,yin2014jstsp,adhikary13tit}.
Notice that the estimation error
due to non-orthogonal pilots in (\ref{ResidualR}) depends on the set
of cyclic shifts: $\{\tau_{l,k}\}$.
Through exploiting the diversity in the covariance matrices of different paths,
we can judiciously choose the set of cyclic shift values:
$\{\tau_{l,k}\}$ in (\ref{CyclicShift}) to minimize the amount of extra
estimation error due to inter-cell interference. This set of optimal cyclic shift
values will align the PDPs of the users with non-orthogonal pilots in a way to
mitigate the inter-cell pilot contamination. In Fig. \ref{ChModelFig},
the composite AoAs of user-$(1,1)$ and user-$(2,1)$ at BS-$2$ are similar and the
existing approaches in \cite{yin2013jsac,yin2015spawc} will not be able to separate
them well. In other words, without alignment, there will be strong pilot
contamination between the two users. However, after aligning path-$1$ to
path-$b$ and path-$2$ to path-$a$, we can expect near interference-free performance
in estimating the channels from user-$(2,1)$ to BS-$2$.

To optimize the overall system performance, we need to solve the following optimal PDP alignment problem:
\begin{eqnarray}
\underset{\{\tau_{l,k}\}}{\textrm{minimize}} & \sum_{b=1}^{B}\sum_{n=1}^N{\tt Tr}({\bm R}_{n}^{(b)}) \label{OptAlignment}\\
  \textrm{subject to} & {\bm P}_{u}^{(b,b)}\tilde{\bm P}_{k}^{(b,b)}={\bm 0}
, \nonumber\\
&b\in[1,B], u\neq k\in[1,K_b], \label{Constraint}
\end{eqnarray}
where the constraint in (\ref{Constraint}) comes from the intra-cell orthogonality
requirement in (\ref{OrthoCond3}).

The PDP alignment problem in (\ref{OptAlignment}) requires exhaustive searches over
all possible cyclic time shifts of all served users. It becomes too complex to be
implemented in practice as the number of served users goes large. Low-complexity
designs are worthwhile and will be discussed in Section \ref{Protocol}.

Summarizing, after aligning different users' PDPs appropriately, we can achieve the
following two benefits at the same time:
\begin{itemize}
  \item 1). For sparse PDPs, more users served by a common BS can transmit
  orthogonal pilot sequences within one OFDM symbol;
  \item 2). When the aligned interfering paths have non-overlapping AoAs with the
  desired path, asymptotic inter-cell interference-free estimation performance can
  be achieved as the size of the massive antenna array goes large, i.e. $M\rightarrow\infty$.
\end{itemize}

\section{Low-Complexity Designs} \label{Protocol}
The optimal solution of the optimization problem in (\ref{OptAlignment}) is
hard to find due to the complex structure of the objective function. Instead, we can
make some simplifications and try to solve easier problems. In the following
discussions, we will assume that the delay spread of users' channels is
$N_{cp}$ chips.

\subsection{Pilot Sequence Length: $N$}\label{Scheme1}
\begin{figure}[t]
\centering \epsfig{file=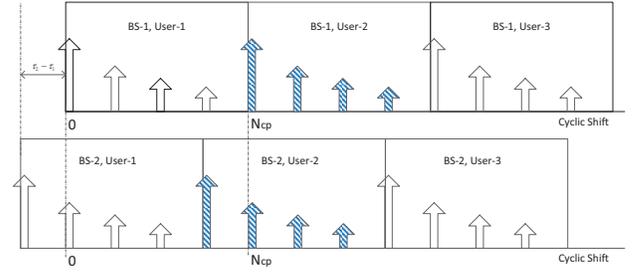,width=0.45\textwidth}
\caption{Low-complexity pilot designs with sequence length equal to $N$.} \label{LengthN}
\vspace{-0.3cm}
\end{figure}

In this case, user-$(l,k)$ will employ the pilot sequence defined in (\ref{CyclicShift}) with $\tau_{l,k}=\tau_l+(k-1)N_{cp}, k=1,...,N/N_{cp}$:
\begin{equation}
{\bm S}_k^{(l)} = {\sf Diag}\left\{1,e^{j\frac{2\pi\tau_{l,k}}{N}},...,e^{j\frac{2\pi\tau_{l,k}(N-1)}{N}}\right\}\cdot{\bm S}_0. \label{CyclicShift2}
\end{equation}
This design, as illustrated in Fig. \ref{LengthN}, will ensure the intra-cell
orthogonality constraint in (\ref{Constraint}) and the optimization problem in
(\ref{OptAlignment}) reduces to:
\begin{eqnarray}
\underset{\{\tau_{l}\}_{l=1}^B}{\textrm{minimize}} & \sum_{l=1}^{B}\sum_{n=1}^N{\tt Tr}({\bm R}_{ n}^{(l)}) \label{OptAlignment2}\\
  \textrm{subject to} & \tau_l\in[0,N-1], l=1,...,B. \label{Constraint2}
\end{eqnarray}
In this simplified problem, we only need to optimize the objective over
$B$ variables: $\{\tau_l\}_{l=1}^B$. Meanwhile, the assignment of the $N/N_{cp}$
orthogonal cyclic shifts to the users served by one BS can also be
optimized.

\subsection{Pilot Sequence Length: $N_{cp}$}\label{Scheme2}
\begin{figure}[t]
\centering \epsfig{file=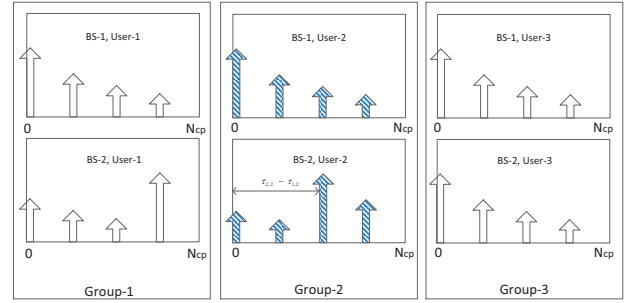,width=0.45\textwidth}
\caption{Low-complexity pilot designs with sequence length equal to $N_{cp}$.} \label{LengthNoverNcp}
\vspace{-0.3cm}
\end{figure}

In this case, user-$(l,k)$ transmits pilots on $N_{cp}$ equally spaced tones:
${\cal G}_k=\{k-1+nN/N_{cp}, n=0,...,N_{cp}-1\}$, $k=1,...,N/N_{cp}$. The pilot sequence is defined as in (\ref{CyclicShift}) but of a shorter length $N_{cp}$:
\begin{equation}
{\bm S}_k^{(l)} = {\sf Diag}\left\{1,e^{j\frac{2\pi\tau_{l,k}}{N_{cp}}},...,
e^{j\frac{2\pi\tau_{l,k}(N_{cp}-1)}{N_{cp}}}\right\}\cdot\tilde{\bm S}_0, \label{CyclicShift3}
\end{equation}
where $\tilde{\bm S}_0$ denotes the length-$N_{cp}$ base sequence.
It is clear that this design will ensure the intra-cell orthogonality
since the users served by one common BS transmit pilots on different sets of tones
(see also Fig. \ref{LengthNoverNcp}). Under the pilot designs in
(\ref{CyclicShift3}), the optimization problem in (\ref{OptAlignment}) can be
decomposed into $N/N_{cp}$ parallel PDP aligning problems: \\
\noindent {\it Sub-problem for ${\cal G}_k$, $k\in[1,N/N_{cp}]$:}
\begin{eqnarray}
\underset{\{\tau_{l,k}\}_{l=1}^B}{\textrm{minimize}} & \sum_{l=1}^{B}\sum_{n=1}^{N_{cp}}{\tt Tr}({\bm R}_{n}^{(l)}) \label{OptAlignment3}\\
  \textrm{subject to} & \tau_{l,k}\in[0,N_{cp}-1], l=1,..,B. \label{Constraint3}
\end{eqnarray}
In each simplified sub-problem for tone group ${\cal G}_k$, we only need to optimize
the objective over $B$ variables: $\{\tau_{l,k}\}_{l=1}^B$. The optimal cyclic
shifts for the interfering users belonging to different tone groups can be derived
independently. Additionally, the allocation of the users served by one BS to the
$N/N_{cp}$ tone groups: $\{{\cal G}_k\}_{k=1}^{N/N_{cp}}$ can also be optimized.

\section{Simulated Performance}\label{Performance}
In this section, we simulate the system as illustrated in Fig. \ref{ChModelFig},
where there are two neighboring cell-edge users creating strong inter-cell
interference. Other simulation parameters and assumptions are listed as follows:
\begin{enumerate}
  \item OFDM configurations\footnote{We follow the numerology in LTE \cite{LTEBook}.}:
  $T=66.67\mu$s ($1/T=15$kHz), $N=128$, $N_{cp}=8$;
  \item Each BS is equipped with a ULA of size $M=50$ and antenna spacing
  $D=\lambda/2$. Each path toward the BS is generated according to the SCM defined
  in \cite{SCM};
  \item Both uniform and exponential PDPs are simulated. For the uniform PDP, we
  have $P_n=P_0,\forall n\in[1,N_{cp}]$. For the exponential PDP, we have $P_n=P_0e^{-0.6(n-1)}$, $\forall n\in[1,N_{cp}]$.
  \item User-$(1,1)$ and user-$(2,1)$ are close to each other but are served by
  different BSs. They share the same scatterers toward each BS, i.e. their
  visibility regions (VR) are overlapping \cite{cost2100};
  \item User-$(1,1)$ and user-$(2,1)$ are assigned to the same tone group as
  defined in Section \ref{Scheme2} and the optimal PDP alignment is found by
  minimizing the cost function in (\ref{OptAlignment3}).
\end{enumerate}

\begin{figure}[t]
\centering \epsfig{file=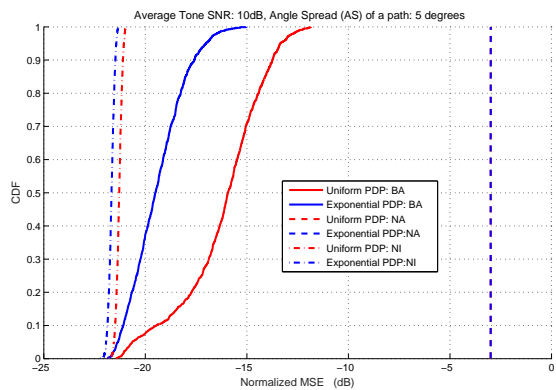,width=0.48\textwidth}
\caption{PDP alignment performance with different PDPs (BA: Best Alignment;
NA: No Alignment; NI: No Interference).} \label{DiffPDP}
\vspace{-0.5cm}
\end{figure}

In Fig. \ref{DiffPDP}, we show the cumulative distribution functions (CDF) of
the normalized mean-square error (MSE)\footnote{
Normalized MSE is defined as: ${\tt NMSE}:=
\frac{\sum_{l=1}^2\sum_{n=1}^{N_{cp}}\|\hat{\bm h}_{n}^{(l,l)}-
{\bm h}_{n}^{(l,l)}\|^2}{\sum_{l=1}^2\sum_{n=1}^{N_{cp}}\|{\bm h}_{n}^{(l,l)}\|^2}$.
} of the estimated channels with and without PDP alignment. In each one of the
$1000$ Monte-Carlo runs, the second-order statistics of the channel taps are
randomly generated according to the SCM. The PDP alignment in (\ref{OptAlignment3})
is based on the second-order statistics to mitigate the pilot contamination.
From Fig. \ref{DiffPDP}, we see the exponential
PDP can provide better decontamination than the uniform PDP. Compared with the case
without alignment, optimal PDP aligning can bring about $13$dB improvement in UL
channel estimation at the median point, i.e. $50\%$ in the CDF curve.

\begin{figure}[t]
\centering \epsfig{file=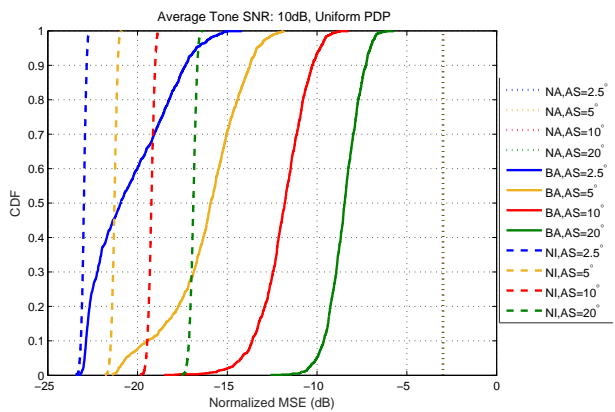,width=0.48\textwidth}
\caption{PDP alignment performance with different ASs.} \label{DiffAS}
\vspace{-0.5cm}
\end{figure}

\begin{figure}[t]
\centering \epsfig{file=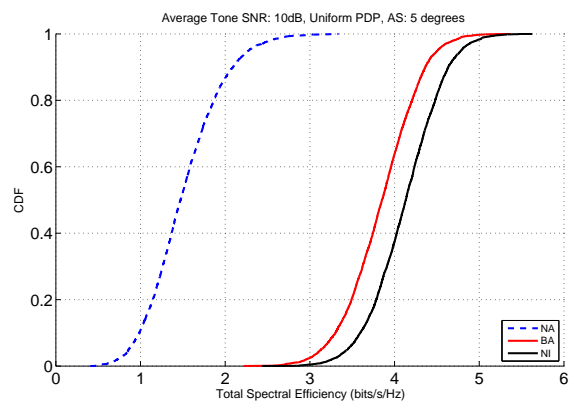,width=0.48\textwidth}
\caption{Achievable sum spectral efficiency with PDP alignment.} \label{SumRateCDF}
\vspace{-0.5cm}
\end{figure}

In Fig. \ref{DiffAS}, we examine the PDP alignment performance for different AS
values. From the plotted curves, we see the PDP alignment favors the scattering
environment with a small AS. In Fig. \ref{SumRateCDF}, we compare the sum of DL
spectral efficiency to the two users in Fig. \ref{ChModelFig} when the BSs
formulate the DL matched-filter precoding vectors with the estimated UL channnels
assuming TDD channel reciprocity. It can be observed that the achieved spectral
efficiency after PDP aligning is pretty close to that without UL pilot contamination.

\vspace{-0.15cm}
\section{Conclusion}\label{conclusion}
\vspace{-0.15cm}
In this paper, relying on a realistic massive MIMO OFDM system model,
we have addressed the issue of pilot contamination and proposed to rely on
PDP aligning to mitigate this type of inter-cell pilot interference.
On one hand, we have shown the UL pilots from the users served by one common BS
can be made orthogonal through PDP aligning. On the other hand, due to the massive
amount of antennas at the BS, we have shown that PDP alignment can also help to
alleviate the pilot contamination thanks to the fact that different paths in
time are originating from different AoAs. Numerical simulations validate that
the pilot contamination can be significantly reduced through aligning the PDPs of
the users served by different BSs appropriately. The proposed PDP aligning can serve
as the new baseline design philosophy for the UL pilots in massive MIMO.

\bibliographystyle{IEEE}

\end{document}